\documentclass[onecolumn,fleqn,usenatbib]{mnras}

\usepackage[T1]{fontenc}
\usepackage{ae,aecompl}
\usepackage{hyperref}

\usepackage{graphicx}    % Including figure files
\usepackage{amsmath}    % Advanced maths commands
\usepackage{amssymb}    % Extra maths symbols

\newcommand{\source}{G\,\ensuremath{17.8+16.7}}

\title[G17.8+16.7: A New Supernova Remnant]{G17.8+16.7: A New Supernova Remnant}

\author[Araya et al.]{
M.~Araya$^{1}$\thanks{E-mail: miguel.araya@ucr.ac.cr},%orcid 0000-0002-0595-9267
N.~Hurley-Walker$^{2}$,
S.~Quir\'os-Araya$^{1}$
\\
$^{1}$Escuela de F\'isica, Universidad de Costa Rica, Montes de Oca, San Jos\'e, Costa Rica, 11501-2060\\
$^{2}$International Centre for Radio Astronomy Research, Curtin University, Kent St, Bentley, WA 6102, Australia}

\date{Accepted 2021 December 02. Received 2021 December 02; in original form 2021 September 12}

\pubyear{2021}

\newcommand{\Fig}{Fig.}

\newcommand{\Sect}{Section}

\begin{document}
\label{firstpage}
\pagerange{\pageref{firstpage}--\pageref{lastpage}}
\maketitle

\begin{abstract}
Non-thermal radio emission is detected in the region of the gamma-ray source FHES\\J$1723.5-0501$. The emission has an approximately circular shape $0.8\degr$ in diameter. The observations confirm its nature as a new supernova remnant, \source. We derive constraints on the source parameters using the radio data and gamma-ray observations of the region. The distance to the object is possibly in the range 1.4--3.5\,kpc. An SNR age of the order of 10\,kyr is compatible with the radio and GeV features, but an older or younger SNR cannot be ruled out. A simple one-zone leptonic model naturally explains the multi-wavelength non-thermal fluxes of the source at its location outside the Galactic plane.
\end{abstract}


\section{Introduction} \label{sec:intro}
Supernova remnants (SNRs) are important for understanding the Galaxy. They heat up the interstellar medium, enrich the environment with heavy elements and accelerate cosmic rays. Almost 300 SNRs are listed in recent catalogs \citep[e.g.,][]{2014BASI...42...47G,2019JApA...40...36G} and more objects and candidates are being found, for example in deeper radio observations \citep[e.g.,][]{2017A&A...605A..58A,2019PASA...36...48H}. However, several thousand SNRs are expected to be found in the Galaxy based on the supernova rate. The discrepancy might be due to selection effects preventing the discovery of very dim SNRs or sources located outside the Galactic plane.

SNRs are known to accelerate particles to relativistic energies. The radiation that these particles produce shows characteristic non-thermal features that can help identify an object as an SNR. Historically, the detection of a synchrotron spectrum in the radio emission of an SNR has been the natural way to confirm the presence of such an object \citep[for a review see, e.g.,][]{2015A&ARv..23....3D}.

High-energy emission from SNRs from X-rays to gamma rays has also been detected and studied in detail \citep{2008ARA&A..46...89R,2016ApJS..224....8A}. Sometimes the accompanying pulsar wind nebula (PWN) associated with an SNR also produces non-thermal emission up to gamma-ray energies. Deep gamma-ray, unbiased, surveys could be useful to find previously unknown SNRs, PWNe or candidates \citep[see, e.g.,][]{2018A&A...612A...1H,Ackermann_2018,2020ApJ...903L..14A,2020ApJ...905...76A}.

The gamma-ray source FHES~J\,$1723.5-0501$ was discovered outside the Galactic plane by \cite{Ackermann_2018} with data from the \emph{Fermi} Large Area Telescope (LAT). They searched for extended GeV sources in high Galactic latitude regions. They noted the presence of an unclassified radio shell in the 1.4 GHz continuum emission data from the NVSS \citep{1998AJ....115.1693C} along the southwestern edge of the somewhat larger source FHES~J\,$1723.5-0501$. This gamma-ray source was therefore proposed to be potentially associated to an SNR or a PWN. In the LAT 4FGL catalog \citep{2020ApJS..247...33A} two gamma-ray sources are found in the region: the extended source 4FGL~J\,$1723.5-0501$e (associated to FHES~J\,$1723.5-0501$), whose energy ($E$) spectrum is described by a simple power-law function ($\frac{dN}{dE}\propto E^{-\gamma}$, with $\gamma$ the spectral index), and the point source 4FGL~J$1722.8-0418$, having a curved spectrum described by a log-parabola ($\frac{dN}{dE}\propto E^{-\alpha - \beta \log{E}}$, with $\alpha$ and $\beta$ some constants). The morphology of this extended source is described by a 2D Gaussian with a 68\%-containment radius of $0.73\degr$.

In this paper we present an analysis of radio observations in the region of FHES~J\,$1723.5-0501$ that confirm the existence of a shell located within the extent of the gamma-ray source. Our analysis reveals a non-thermal spectrum for the radio emission. The shell-like appearance and spectrum establish this object as a new SNR, labeled \source{}. We also analyzed gamma-ray data from the LAT which confirm that the emission in the region is best described by an extended source with a hard GeV spectrum. In \Sect~\ref{sec:data} we describe the radio and GeV data analyses, and in \Sect~\ref{discussion} we use the properties of the emission to constrain some parameters such as the source distance and age.

\section{Data analysis}\label{sec:data}

\subsection{Radio observations and data analysis}\label{sec:radio}

\begin{figure}
    \includegraphics[width=\textwidth]{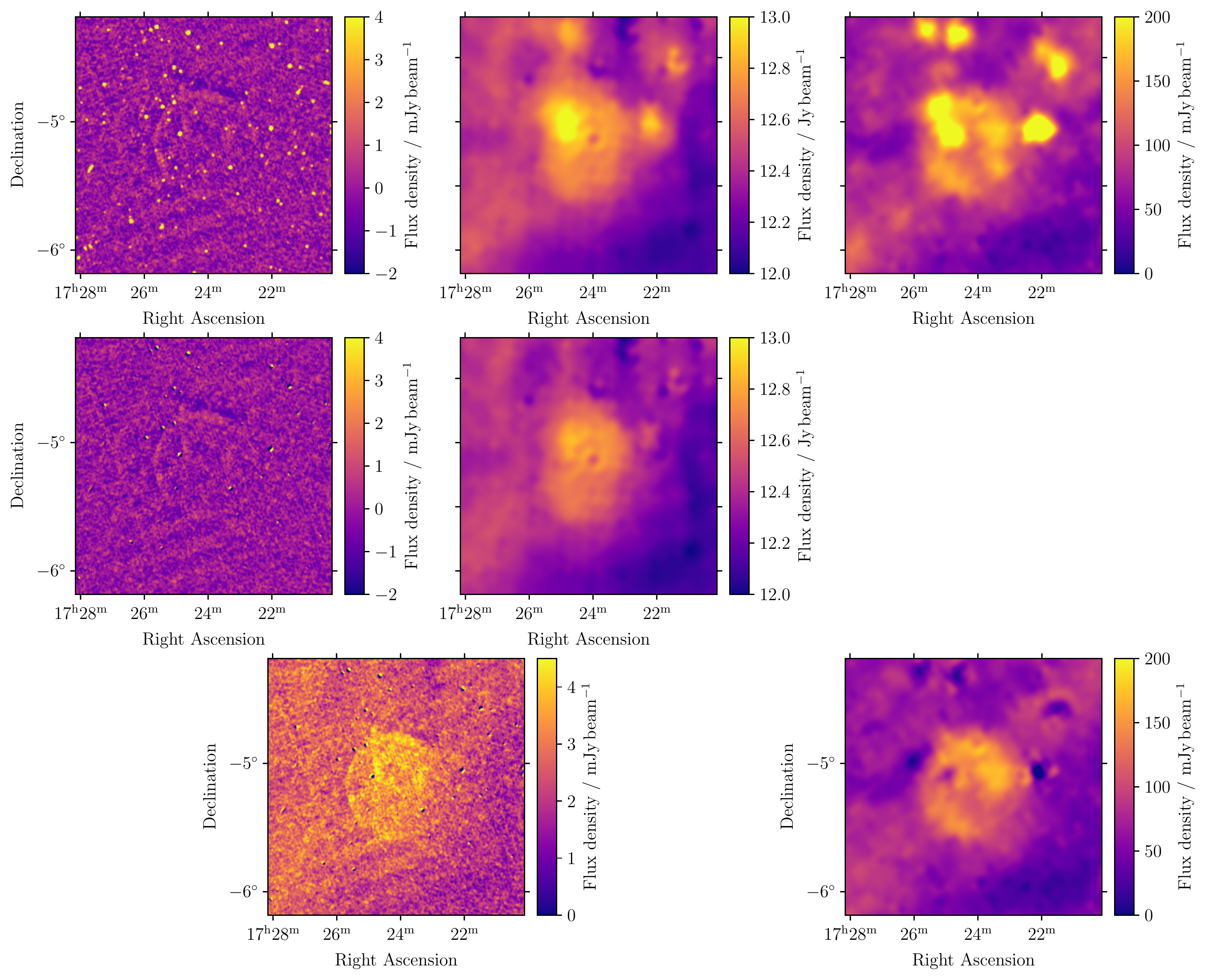}
    \caption{The radio data of this region as processed in \Sect~\ref{sec:radio}. From left to right, the top row shows raw data from NVSS, CHIPASS, and SPASS. The second row shows the NVSS (left) and CHIPASS (right) data after source-subtraction. The bottom row shows the result of feathering the NVSS and CHIPASS data together (left) and the source-subtracted SPASS data (right).}
    \label{fig:all_radio}
\end{figure}

Given the high Galactic latitude of \source{}, it can only be detected in relatively shallow wide-area sky surveys, rather than the deeper images often available from targeted Galactic plane surveys. Of the archives we searched, it was most clearly detected in the NRAO Very Large Array Sky Survey \citep[NVSS; ][]{1998AJ....115.1693C} as a partially-resolved crescent of diameter $\sim0.95\degr$ (top-left panel of \Fig~\ref{fig:all_radio}, which also shows other data used in this section). This is larger than the maximum angular scale to which NVSS is sensitive, so a total flux density cannot be measured from these data alone. To fill in these larger angular scales, we used the Continuum map of the HI Parkes All-Sky Survey \citep[CHIPASS; ][]{2014PASA...31....7C}, a radio sky survey\footnote{\href{https://www.atnf.csiro.au/people/mcalabre/CHIPASS/index.html}{https://www.atnf.csiro.au/people/mcalabre/CHIPASS/index.html}} at 1.4\,GHz covering Dec~$<+25\degr$ at $14.4'$ resolution. Following the method of \cite{2021A&A...648A..30B}, we selected all compact sources detected in NVSS in the region, convolved them to match the CHIPASS resolution, produced an output FITS image in the same sky frame as the CHIPASS data, and subtracted the NVSS model from the CHIPASS image (in Jy\,beam$^{-1}$).

We used the software \textsc{poly\_flux} \citep{2019PASA...36...48H} to measure the total flux 
densities of \source{} in each band; the tool estimates and subtracts a mean background level. Since the 
selection of the boundaries of the SNR is somewhat subjective, we used the tool ten times and 
recorded the average result, finding that the total flux density at 1.4\,GHz is $2.1\pm0.1$\,Jy. 

\begin{figure}
    \includegraphics[width=\textwidth]{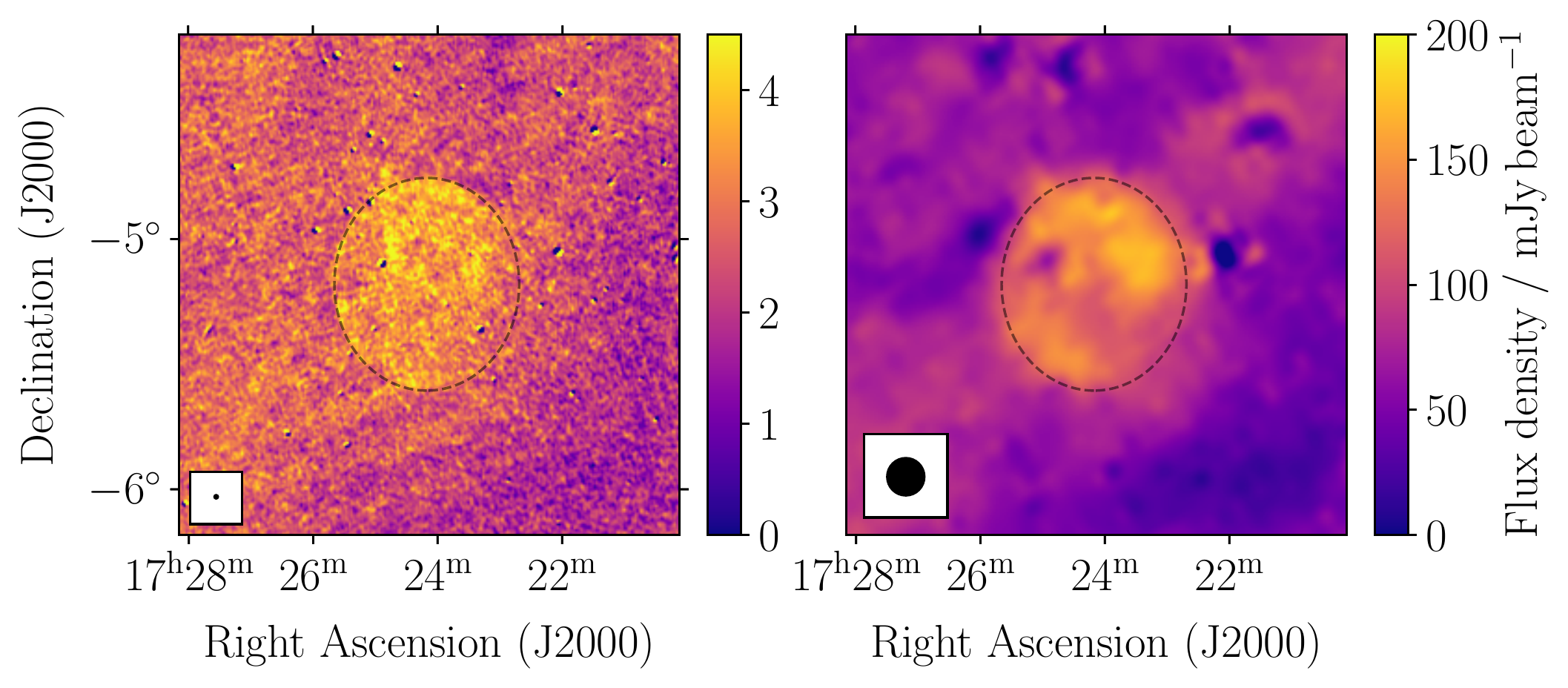}
    \caption{Four\,deg$^2$ of the region surrounding \source{} as seen at 1.4\,GHz (left) and 2.3\,GHz (right) after source subtraction as described in \Sect~\ref{sec:radio}; the subtracted CHIPASS and NVSS data (left) have been feathered together to produce an image to properly capture angular scales $>45''$. Global background levels of 31\,mJy\,beam$^{-1}$ and 640\,mJy\,beam$^{-1}$ have been subtracted from the left and right images, respectively. An ellipse centred on $17^\mathrm{h}24^\mathrm{m}10.5^\mathrm{s} -5\degr10'52.5''$, size $51'\times45'$, and orientation due North, is shown by a dashed black semi-transparent line on both panels. The full-width-half-maximum of the point spread function is shown as a solid black ellipse in the lower-left of each panel.}
    \label{fig:radio}
\end{figure}

We performed a similar process on the S-Band Polarization All Sky Survey \citep[SPASS ;][]{2019MNRAS.489.2330C}, a 2.3-GHz survey of polarized radio emission covering the Southern sky (Dec~$<-1\degr$) at $8.9'$ resolution. To scale the flux densities of the NVSS sources from 1.4\,GHz to 2.3\,GHz, we used the spectral indices from the catalogue produced by \cite{2018MNRAS.474.5008D}.
Subtracting this model from the S-PASS Stokes~I results in the right-hand panel of \Fig~\ref{fig:radio}. 
Running \textsc{poly\_flux} repeatedly we find that the total flux density at 2.3\,GHz is $1.45\pm0.05$\,Jy; 
the uncertainty is dominated by the less clean source subtraction, particularly for two sources on the edge of the shell. 

Since CHIPASS and NVSS sample different but complementary angular scales, it is instructive to combine the images in the Fourier domain, a process known as ``feathering'' \citep{2017PASP..129i4501C}. We used a custom \textsc{python} command\footnote{\href{https://github.com/nhurleywalker/feather}{https://github.com/nhurleywalker/feather}} derived from the implementation in the Common Astronomy Software Applications (\textsc{CASA}) to perform this operation, and the result is shown in the left panel of \Fig~\ref{fig:radio}. \source{} is clearly visible as a sharp-edged elliptical shell which is quite filled, and brighter and more defined toward the Eastern side.

No detections with signal-to-noise sufficient to determine a radio flux density were made in any other sky surveys. From the 1.4 and 2.3\,GHz flux density measurements we calculated a two-point spectral index, finding $\alpha=-0.75\pm 0.15$ for $S\propto\nu^\alpha$. This is consistent with non-thermal emission from a synchrotron-emitting shell supernova remnant.

\subsection{Archival X-ray observations}\label{sec:xray}

\begin{figure} 
    \includegraphics[width=\textwidth]{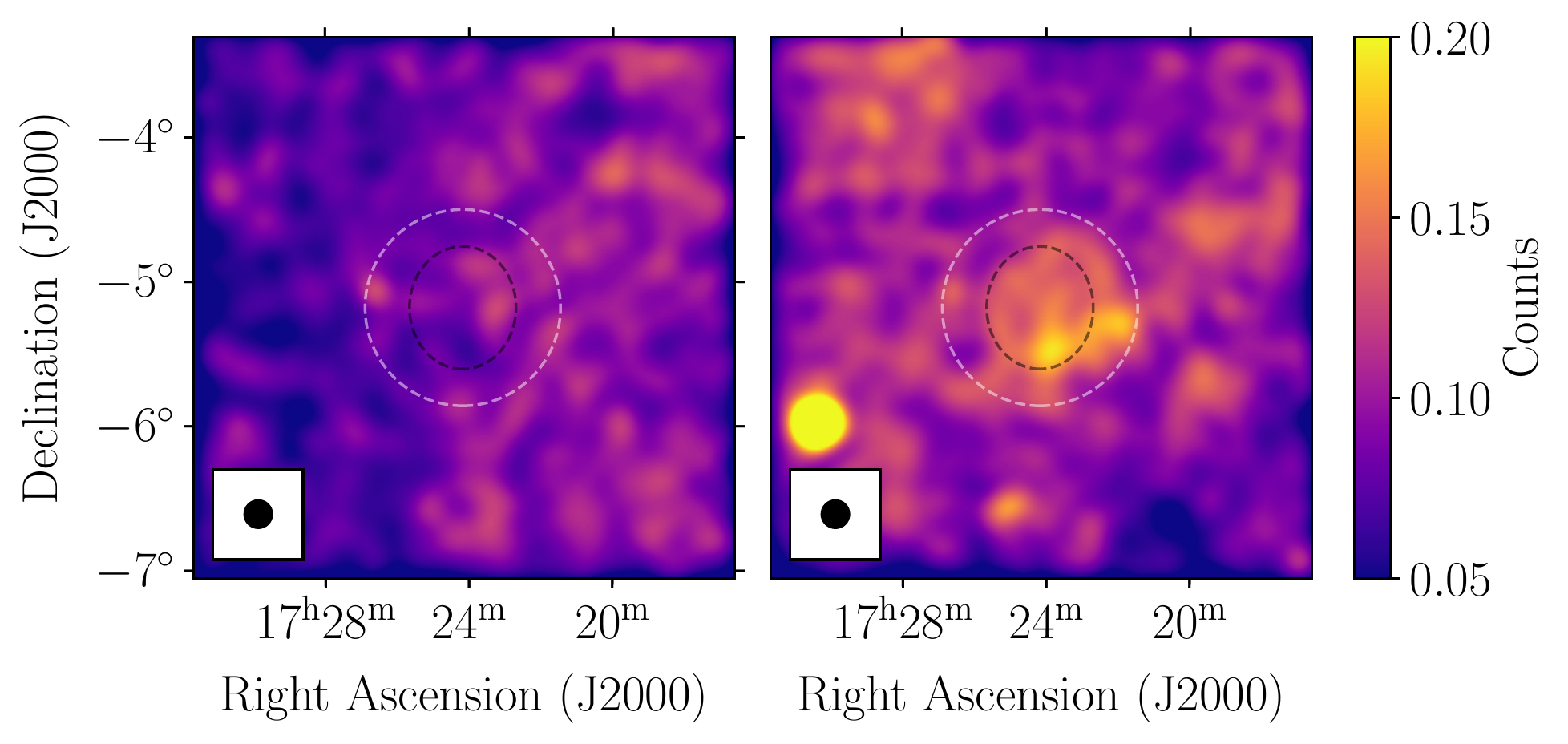}
    \caption{13\,deg$^2$ of the region surrounding \source{} as seen by the RASS (\Sect~\ref{sec:xray}); the left panel shows the soft band (0.1 -- 0.4\,keV) and the right panel shows the hard band (0.5 -- 2.0\,keV). Both images have been convolved by a Gaussian kernel with full-width-half-maximum $675''$ (shown as a solid black ellipse in the lower-left of each panel). The radio shell is indicated by the same ellipse as in \Fig~\ref{fig:radio}, and the best-fitting spatial model to the gamma-ray data (Gaussian + PS; see \Sect~\ref{sec:gamma}) is shown as a white dashed ellipse.}
    \label{fig:xray}
\end{figure}

We searched the archives of the X-ray observatories but found no pointed (deep) observations that covered this field. On initial inspection, \source{} is not visible in the ROSAT All-Sky Survey \citep[RASS;][]{1999A&A...349..389V}. However, after convolution with an $\sim11'$ Gaussian kernel, some emission is visible in the RASS hard-energy band (0.5 -- 2.0\,keV); there is no detection in the soft-energy band (0.1 -- 0.4\,keV; see \Fig~\ref{fig:xray}). The hard X-ray emission correlates well with the radio shell on the Eastern side, and correlates more clearly with the gamma-ray emission (\Sect~\ref{sec:gamma}) on the Western side. The counts within the radio shell are about 50\,\% higher than the background in this area.

\subsection{Gamma-ray observations}\label{sec:gamma}
The \emph{Fermi}-LAT is a converter/tracker telescope detecting gamma rays in the energy range between 20 MeV and $\ga$1 TeV \citep{2009ApJ...697.1071A}. We gathered LAT data taken from August~2008 to July~2021 in the energy range 0.5--500\,GeV for the analysis. We used the software {\tt fermitools} version~2.0.8 by means of the {\tt fermipy} package version~1.0.0 to perform the analysis. We applied the recommended cuts for the analysis\footnote{See https://fermi.gsfc.nasa.gov/ssc/data/analysis/documentation/Cicerone/Cicerone\_Data\_Exploration/Data\_preparation.html}, selecting good quality front and back-converted events in the {\tt SOURCE} class ({\tt evclass=128}, {\tt evtype=3}), and having zenith angles lower than $90\degr$ to avoid contamination from Earth's limb. We used the corresponding response functions for {\tt Pass 8} analysis, {\tt P8R3\_SOURCE\_V3} and binned events with a spatial scale of $0.05\degr$ and with ten bins per decade in energy for exposure calculations. We included events reconstructed within $15\degr$ from the coordinates RA (J2000) $=17^\mathrm{h}24^\mathrm{m}00^\mathrm{s}$, Dec (J2000) $=-5\degr12'00''$, defined as the region of interest (RoI). We included the sources in the 4FGL-DR2 catalog \citep{2020ApJS..247...33A,2020arXiv200511208B} that are located within $20\degr$ of the center of the RoI, except for 4FGL~J\,$1723.5-0501$e and 4FGL~J\,$1722.8-0418$, since we carried out a more detailed analysis of their emission. We modeled the diffuse Galactic emission and the isotropic emission (including the residual cosmic-ray background) with the standard files {\tt gll\_iem\_v07.fits} and {\tt iso\_P8R3\_SOURCE\_V3\_v1.txt}, respectively, provided by the LAT team\footnote{See \url{https://fermi.gsfc.nasa.gov/ssc/data/access/lat/BackgroundModels.html}}. We applied the energy dispersion correction to all sources except for the isotropic diffuse emission component, as recommended by the LAT team\footnote{See https://fermi.gsfc.nasa.gov/ssc/data/analysis/documentation/Pass8\_edisp\_usage.html}. We applied the maximum likelihood technique \citep{1996ApJ...461..396M} and fitted any free morphological or spectral parameters of the sources in order to maximize the probability for the model to explain the data. The detection significance of a source can be calculated using the test statistic (TS), defined as $-2\log(\mathcal{L}_0/\mathcal{L})$, with $\mathcal{L}$ and $\mathcal{L}_0$ being the maximum likelihood functions for a model containing the source and for the model without the additional source (the null hypothesis), respectively.

\begin{figure}
    \centering
    \includegraphics[width=0.7\textwidth]{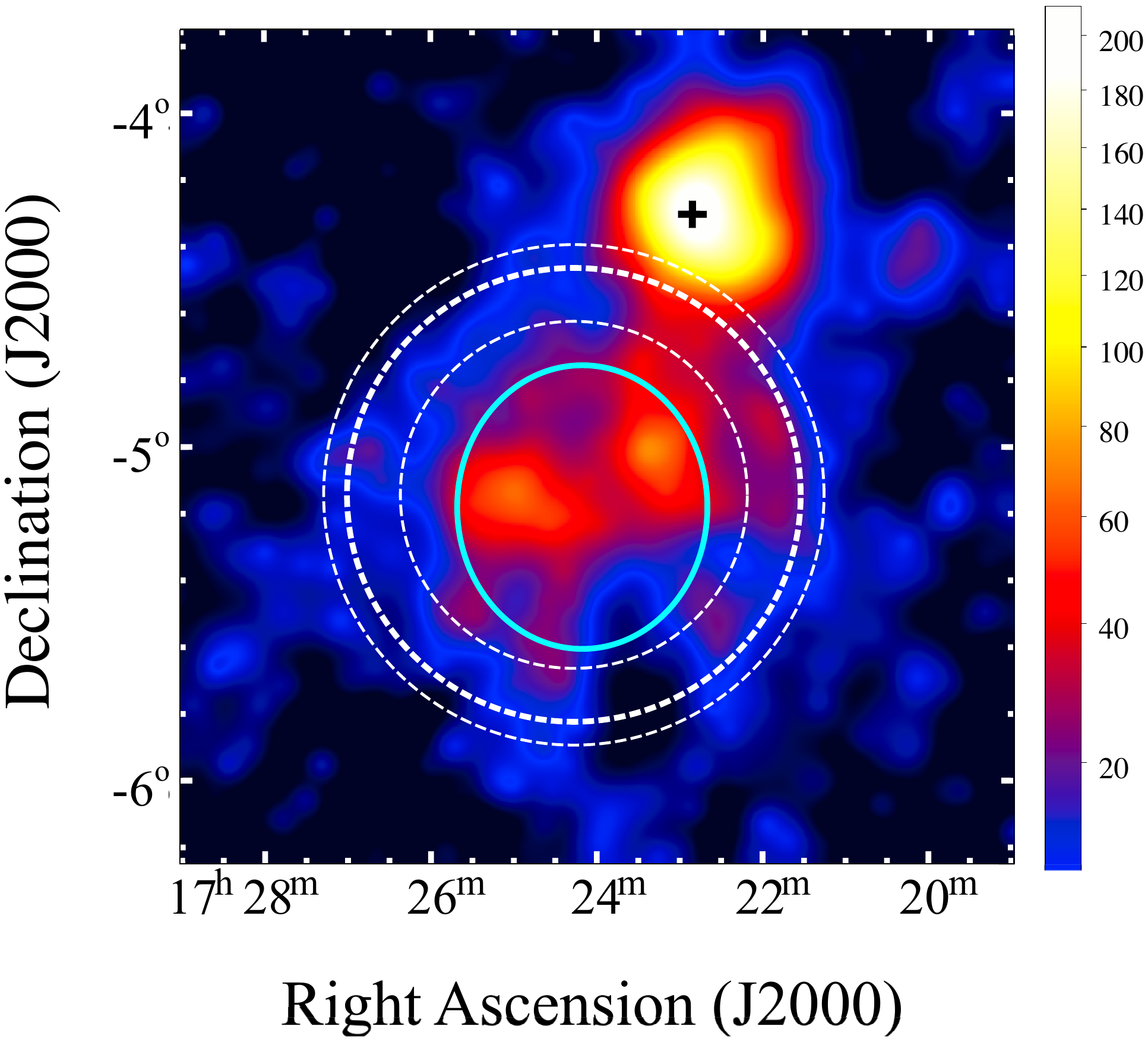}
    \caption{TS map obtained with LAT data in the energy range 1--500\,GeV. The image is obtained by moving a putative point source through each pixel in the map and calculating its TS value. The solid cyan ellipse corresponds to the radio shell ellipse shown in \Fig~\ref{fig:radio}. The circles represent the 68\%-containment radius of the 2D Gaussian found in our analysis (thick dashed line) and its $1\sigma$ uncertainty region (thin dashed lines). The location of the point source 4FGL~J\,$1722.8-0418$ is indicated by the cross.}
    \label{fig:tsmap}
\end{figure}

Taking advantage of the improved point-spread function of the LAT at higher energies, we performed a morphological analysis of the emission in the region of \source{} using events with reconstructed energies above 5\,GeV. We left free the spectral normalizations of the 4FGL sources located within $10\degr$ of the RoI center, while for the sources located within $5\degr$ of the RoI center we left all spectral parameters free. We used the Akaike Information Criterion \citep{1974ITAC...19..716A}, defined as AIC = $2k - 2\ln(\mathcal{L})$, where $k$ is the number of free parameters in the model, to compare the relative quality of the morphological models. We compared the following hypotheses for the emission: a symmetric 2D Gaussian, a uniform disk, a symmetric 2D Gaussian with a point source and a uniform disk with a point source \citep[for the definition of the extended models, see][]{2012ApJ...756....5L}. For each case we let the location and extension of the sources free to vary and performed a likelihood profile scan to find the maximum model likelihood. We also assumed simple power-laws for the spectra of the sources, which is justified below. The results are shown in Table \ref{table:LAT}. A model consisting of a Gaussian and a point source result in the lowest AIC value and is therefore the best description among the tested models. The point source was found with the tool {\tt find\_sources} and its location is consistent with that of 4FGL J1722.8--0418, seen outside the radio shell of \source{} (see \Fig~\ref{fig:tsmap}). Including the point source 4FGL~J\,$1722.8-0418$ in the model improves the quality of the fit and is important to correctly estimate the size of the extended source, as seen in Table \ref{table:LAT}. We quantified the significance of extension by calculating twice the difference between the $\log \mathcal{L}$ for an extended source model and that obtained with a point-like source model at its best-fit position. In all cases the extended source model is preferred over a point source for the emission in the region of \source{}. For the last two models in Table \ref{table:LAT} we found TS$_{\mbox{\tiny ext}} = 55.3$ for the disk model and TS$_{\mbox{\tiny ext}} = 65.6$ for the Gaussian. The 68\%-containment extension of the Gaussian in our best-fit model is consistent with the extension reported for 4FGL~J\,$1723.5-0501$e in the 4FGL catalog \citep{2020ApJS..247...33A}.

We estimated the effect of the systematic uncertainty in the model of the diffuse Galactic emission on the measurement of the source extension. The uncertainties in this model could be important for the treatment of extended sources, even though the effect is expected to be larger at the lowest energies where this emission dominates. We used the eight alternative model files developed originally by \cite{2016ApJS..224....8A}, scaled appropriately to account for differences in energy dispersion between {\tt Pass 7} and {\tt Pass 8} reprocessed data\footnote{See https://fermi.gsfc.nasa.gov/ssc/data/access/lat/Model\_details/Pass8\_rescaled\_model.html}. We fitted the source extension using both the uniform disk and the 2D Gaussian for each alternative Galactic diffuse emission model and estimated the systematic uncertainty as in \cite{2016ApJS..224....8A}. The systematic uncertainty for the 68\%-containment radius of the Gaussian is $0.15\degr$, and that of the disk radius $0.02\degr$.

We also searched for hints of energy dependent morphology, as expected for example from electron cooling and transport in PWNe. We divided the data in two energy intervals, 5--40\,GeV and 40--500\,GeV. These energy intervals were chosen as to contain enough statistics to confirm a significant extension of the source with TS$_{\mbox{\tiny ext}} > 20$. We fitted the source extension in each interval and the results for the low and high-energy data sets were $0.54^{+0.14}_{-0.10}\degr$ and $0.92^{+0.22}_{-0.17}\degr$ for the 68\%-containment radii of the 2D Gaussian, and $0.54^{+0.04}_{-0.03}\degr$ and $0.66^{+0.05}_{-0.04}\degr$ for the disk radii, respectively ($1\sigma$ statistical errors given). The discrepancy between the measured extensions at the low and high energies, considering statistical uncertainties only, is then at the $1.5-2\sigma$ level, depending on the morphological model adopted. Slightly higher extensions are seen at higher energies, which is the opposite behaviour to that expected in a PWN. The centroid locations found for the disks at the low and high energies are incompatible at the $3\sigma$ level, considering their statistical uncertainties only, while the $1\sigma$ error ellipses on the location of the centroids found with the Gaussian templates do overlap. The centroids of the emission at the highest energies are shifted towards the south east with respect to those found in the low energy interval.

In order to further study this slight tension we performed a new fit in the 5--500\,GeV energy range to probe the existence of two different extended sources. We modeled the GeV emission from the SNR with a uniform disk with a fixed radius of $0.4\degr$, centred at the position described in \Fig~\ref{fig:radio}, and we searched for an additional source fitting its location and extension by means of a likelihood profile scan under the uniform disk hypothesis. The best-fit radius of the new disk and its $1\sigma$ statistical uncertainty is $1.32^{+0.05}_{-0.06}\degr$, with its centre located within the radio shell of the SNR. This model shows a $\Delta$AIC$=1.6$ with respect to the best-fit model found before including only one extended source, and therefore results in no statistical improvement. Future studies with more statistics will be important to better understand the morphology of the emission.

A TS map obtained from LAT data in a broader energy range, 1--500\,GeV, to improve the statistics, is shown in Fig. \ref{fig:tsmap}. The emission in the region of \source{} as well as that of the point source 4FGL J1722.8--0418 are clearly visible. The GeV emission is more significant within the shell boundary of \source{}, as seen in the figure.

Using events in the entire energy range, 0.5--500\,GeV, we tested for spectral curvature with a log-parabola function. The TS values only marginally improved with respect to the fits using the simple power-law as the spectral shape for 4FGL~J\,$1722.8-0418$ ($\Delta$TS$=2.4$) and for the extended emission ($\Delta$TS$=2.7$), indicating that in this energy range a simple power-law shape is preferred for both sources. The resulting spectral indices (and their $1\sigma$ uncertainties) are $2.44\pm 0.02_{\mbox{\tiny stat}} \pm 0.07_{\mbox{\tiny sys}}$ for 4FGL~J\,$1722.8-0418$ (which has an overall TS$=231.6$), in agreement with the studies of this point source by \cite{2016RAA....16...97D}, and $1.83\pm0.02_{\mbox{\tiny stat}} \pm 0.05_{\mbox{\tiny sys}}$ for the extended source (with TS$=153.2$). The systematic uncertainty in the spectral index for the extended source includes the effects of using the alternative diffuse Galactic background models described earlier, as well as the effect of replacing the Gaussian morphology with the best-fit disk morphology. Changing the morphology to describe the emission is the dominant source of systematic uncertainty. The GeV spectrum of 4FGL~J\,$1722.8-0418$ is clearly softer than that of the extended source. The spectral energy distribution (SED) fluxes of \source{} were obtained dividing the data in ten energy intervals and fitting the normalization in each bin. They are shown below in Fig. \ref{fig:SED}.

\begin{table*}
\caption{Results of the morphological analysis of \emph{Fermi}-LAT data.}
\label{table:LAT}
\begin{center}
\begin{tabular}{lccc}
\hline
\hline
Spatial model & Fitted size$^{a}$ ($\degr$) & $\Delta$AIC$^b$ \\
\hline
Disk & $1.09^{+0.04}_{-0.05}$ & 54.3\\
Gaussian & $0.83^{+0.08}_{-0.07}$ &  36.2 \\
Disk$+$ PS & $0.55^{+0.05}_{-0.03}$ &  10.3\\
Gaussian$+$PS & $0.68^{+0.07}_{-0.16}$ & 0 \\
\hline
\end{tabular}\\
\textsuperscript{$a$}\footnotesize{Radius for the disk and 68\%-containment radius for the Gaussian and their $1\sigma$ statistical uncertainties.}\\
\textsuperscript{$b$}\footnotesize{$\Delta$AIC is equal to the value of AIC for each model minus the AIC value for the best-fit model.}\\
\end{center}
\end{table*}

\section{Discussion}\label{discussion}

\subsection{Limits on age and distance}

In the absence of a measured distance, we can use the morphological and brightness properties of the SNR to infer limits on the physical characteristics. Studies of the Local Group galaxies and Magellanic Clouds have demonstrated that SNR 1.4-GHz luminosities 
typically have values in the range $5\times10^{14} < L_\mathrm{1.4GHz} < 10^{17}$\,W\,Hz$^{-1}$ \citep[e.g.][]{1998ApJ...504..761C}. 
Assuming that \source{} is more luminous than $5\times10^{14}$\,W\,Hz$^{-1}$, we can obtain a limit on its distance 
from Earth by $\sqrt{\frac{L_\mathrm{1.4GHz}}{4\pi S_\mathrm{1.4GHz}}}$, i.e. $d>1.4$\,kpc (and diameter $D>20$\,pc). Assuming a low ISM density of 0.1\,cm$^{-3}$ and using otherwise standard values in the SNR evolutionary model calculator provided by \cite{2017AJ....153..239L}, we find that for $D>20$\,pc, the source must be $>10$\,kyr old. Given its clearly defined edges and filled disk, we suggest it is still in the Sedov-Taylor phase, and not likely to be more than an order of magnitude older than this, which yields $D<50$\,pc and $d<3.5$\,kpc. However, since there are few known high-latitude SNRs, these values are only estimates, and a direct measurement may difficult for faint SNRs without obvious molecular cloud or pulsar associations.

\subsection{Gamma-ray properties}

\begin{figure}
    \includegraphics[width=\textwidth]{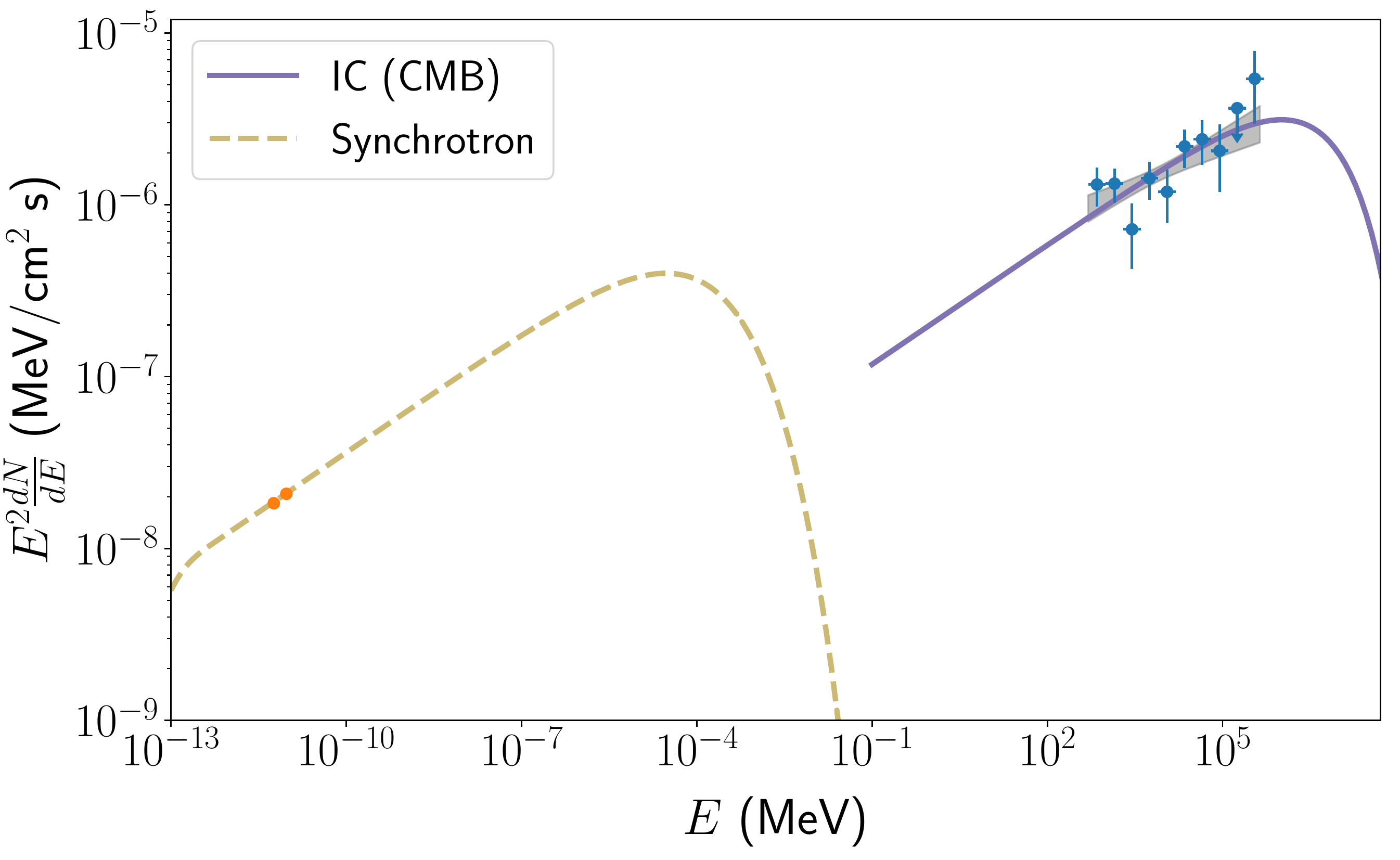}
    \caption{SED of \source{} and the resulting fit to the radio and GeV data under a one-zone IC-CMB scenario. The shaded region represents the propagated $1\sigma$ statistical uncertainty on the spectral fit in the 0.5--500\,GeV energy range.}
    \label{fig:SED}
\end{figure}

A hard GeV spectrum such as that observed for \source is expected under an inverse Compton (IC) scenario, where high-energy electrons accelerated in the shock of the SNR interact with soft ambient photon fields, such as the cosmic microwave background, to produce the gamma rays. This mechanism for the production of high-energy emission would be more natural in a region located outside the Galactic plane where \source{} is found and the matter density is expected to be low. This leptonic-IC scenario for the gamma rays is consistent with our radio observations. Under this model, and given the measured GeV spectral index $\gamma \sim 1.83$ ($\frac{dN}{dE}\propto E^{-\gamma}$), the predicted radio spectral index for the synchrotron emission ($S \propto \nu^{\alpha}$) from the same uncooled population of electrons is $\alpha = 1-\gamma = -0.83$, fully consistent with our measured value of $-0.75\pm 0.15$. We used the {\tt naima} package \citep{naima} to fit the radio and GeV fluxes with a one-zone leptonic model using a particle distribution that is a power-law with an exponential cutoff. The radio fluxes result from synchrotron emission from electrons in a magnetic field while the gamma rays are from IC scattering of cosmic microwave background (CMB) photons by the same electrons. The results are shown in Fig. \ref{fig:SED}. The required magnetic field is $B=1.06^{+0.33}_{-0.23}\,\mu$G, the spectral index and cutoff energy of the lepton distribution are $2.5\pm0.1$ and $57^{+61}_{-30}$ TeV, respectively. The total energy content in the relativistic electrons (integrated above a particle energy of 1\,GeV) amounts to $(1.4^{+0.9}_{-0.6})\times 10^{49}\,\left( \frac{d}{3\, \mbox{\tiny kpc}} \right)^2$\,erg, normalised to an arbitrary distance of 3\,kpc. This is only between 0.3\% and 1.8\% of the typical kinetic energy available in an SNR shock ($10^{51}\,$erg) for a distance range of 1.4--3.5\,kpc. Given that the GeV spectrum of \source{} is described by a power-law with no apparent cutoff, the resulting cutoff in the particle distribution is not well-constrained. Observations in the TeV energy range should be carried out for this purpose.

Another mechanism for the origin of gamma-ray emission are inelastic collisions of relativistic protons, accelerated by the SNR, with ambient protons. In this case a high density of matter is required to enhance the flux of gamma rays. Since the gamma-ray distribution approximately follows the parent proton distribution in this hadronic scenario \citep[see, e.g.,][ for a review]{2013A&ARv..21...70B}, a proton distribution harder than predicted by linear diffusive shock acceleration theory would be required in order to explain the GeV spectrum of \source. However, if dense clumps of gas exist within the shell of an SNR and the highest energy protons are able to interact with this gas, it has been shown that a hard GeV spectrum could be produced \citep{2014MNRAS.445L..70G}.

Our estimated gamma-ray luminosity of \source{} in the 1--100\,GeV energy interval is $\sim9\times 10^{33}$\,erg\,s$^{-1}$ for an arbitrary source distance of 3\,kpc chosen within our estimated range. This luminosity is between one and two orders of magnitude lower than the typical luminosities of evolved SNRs interacting with dense gas clouds \citep{2016ApJS..224....8A}. Our measurements are therefore more consistent with a low-density environment around the SNR and a leptonic scenario for the gamma rays.

The spectral index that we measured for the high-energy emission ($\sim 1.83$) could suggest something regarding the SNR age, for example following the trend seen by \cite{2016ApJS..224....8A} where sources with harder GeV spectra tend to be usually younger SNRs. However, we note that there is a large scatter in the age of their sample SNRs even for hard GeV spectra ($\sim\,$1--60\,kyr). This could be due to the complexity in the environments of SNRs. SNRs with an age $\sim0.5\times 10^4$ yr in low-density environments (with number density $\lesssim 0.1$~cm$^{-3}$) and low magnetic fields are predicted to show gamma-ray fluxes dominated by IC emission with a hard GeV spectrum \citep{2012MNRAS.427...91O,2019ApJ...876...27Y}. Other SNRs with a very similar gamma-ray spectrum are the faint radio source G\,$279.0+1.1$ \citep{2020MNRAS.492.5980A}, whose age is unknown but shows features of an evolved SNR, and a dim SNR recently discovered, G\,$150.3+4.5$, \citep{2014A&A...567A..59G,2020A&A...643A..28D}. Other extended sources with similar GeV spectra, although with no known counterparts at other wavelengths, are G\,350.6--4.7 \citep{2018MNRAS.474..102A} and 2HWC\,J2006+341 \citep{2020ApJ...903L..14A}. Their GeV spectra are similar to those of some young SNRs \citep[SN 1006, RX J1713.7--3946, RCW 86, with ages of 1--2\,kyr,][]{2019PASJ...71...77X,2011ApJ...734...28A,2016ApJ...819...98A}. Another feature of \source{} that resembles a young SNR is its steep radio spectrum \citep[see][for examples of radio spectra]{2014Ap&SS.354..541U}. However, \source{} has fainter X-ray emission than young SNRs (\Sect~\ref{sec:xray}); deeper X-ray observations in this band would yield useful results, e.g. emission lines to help discriminate the progenitor, the properties of the plasma, the distance to the source and the possible presence of non thermal emission from the highest energy electrons.
As pointed out before recent simulations predict the SED of SNRs with ages of up to at least $5\,$kyr to be dominated by IC emission at gamma-ray energies with low synchrotron fluxes \citep{2019ApJ...876...27Y}. Such an age is more compatible with the predicted value ($10^4$\,yr) for a low density environment mentioned earlier.

It is interesting to compare the extension of the radio shell with that of the GeV emission. As seen in Fig. \ref{fig:tsmap} the gamma-ray source is more extended than the radio source. Gamma-ray emitting SNRs show very similar radio and GeV angular diameters \citep{2016ApJS..224....8A}. GeV diameters of LAT-detected SNRs are found within $\sim0.3\degr$ and $\sim$20\% of their radio diameters \citep{2016ApJS..224....8A}. Taking the radius of the GeV disk in our fit as a measure of the radius of the gamma-ray source and its corresponding uncertainty, which results from combining the statistical uncertainty (reported in Table \ref{table:LAT}) with the systematic uncertainty in quadrature, the diameter of the gamma-ray source is in the range 1.03--1.21$\degr$, which is greater than the $\sim$0.8$\degr$-radio diameter by 0.23--0.41$\degr$ (29--51\% larger). Future observations with more statistics might reveal that the discrepancy could be similar to that observed for other SNRs, or that the sizes could indeed be considerably different. There could be several reasons for this. First, deeper radio observations could reveal a larger shell. This possibility has been proposed by \cite{2020MNRAS.492.5980A} to explain a discrepancy for the SNR G\,$279.0+1.1$ GeV and radio extensions. However, G\,$279.0+1.1$ shows an incomplete shell while for \source{} it seems well-defined. Another possibility is the diffusion of relativistic particles that produce gamma rays in the interstellar medium around the SNR. Recent simulations show that electrons could produce a halo of hard GeV emission around an SNR, although having a flux of about 20 to 30\% of the SNR flux \citep{2021arXiv210810773B}, which would make a detection difficult. This scenario should be studied in the future with more data. Another possibility that would account for a discrepancy between the GeV and radio sizes is that part or all of the gamma-ray emission comes from a PWN that is perhaps associated to \source{}, while the radio emission is from the SNR. Gamma-ray emission from PWNe can be quite extended. However, no pulsars are reported in The Australia Telescope National Facility Pulsar Catalogue \citep{2005AJ....129.1993M} within $2.3\degr$ of the center of the SNR, and no composite SNRs are known for which the PWN is larger than the radio shell. Finally, we cannot rule out that a different gamma-ray source lies in the same line of sight as \source{}. We have found a slight discrepancy between the fitted GeV sizes at low and high energies (considering only their statistical uncertainties) which could point towards this latter scenario. Future studies with more statistics and additional multi-wavelength observations will be important to address this issue.

The gamma-ray point source 4FGL~J\,$1722.8-0418$ shows a soft GeV spectrum. The spectrum of LAT pulsars is also usually soft, described by a power-law with an exponential cut-off. The spectral indices are typically found in the range 1--2 and the cut-off energies are of a few GeV \citep{Abdo_2013}. This is not the case for 4FGL~J\,$1722.8-0418$, for which we find a simple power-law spectrum in the 0.5--500 GeV energy range with an index of $\sim 2.44$. However, using  classification algorithms, \cite{2016ApJ...820....8S} found that this source could be a pulsar (either a young or a millisecond pulsar) and it is likely not associated to an active galactic nucleus. No radio emission was found by \cite{2021ApJ...914...42B} at the location of the point source. Assuming 4FGL~J\,$1722.8-0418$ is a pulsar associated with \source{} that has travelled away from the center of the SNR to its current location $0.94\degr$ away, the required pulsar transverse velocity is $2200$\,km\,s$^{-1}\,\left(\frac{d}{1.4\mbox{\tiny\,kpc}} \right)\,\left(\frac{t}{10\mbox{\tiny\,kyr}} \right)^{-1}$, with $t$ the time since the explosion. The mean transverse velocity of pulsars is $\sim 500\,$km\,s$^{-1}$ \citep{1997MNRAS.289..592L}, rarely exceeding $1000\,$km\,s$^{-1}$, although such a high velocity is not impossible. Therefore, for a high transverse velocity of $1000\,$km\,s$^{-1}$ and our distance range of 1.4--3.5\,kpc estimated earlier, an age for the SNR in the range 22--55\,kyr is required. If, on the other hand, no pulsar association is found for \source, it would be consistent with it being the remnant of a type Ia supernova, while most pulsars are expected to be born in the Galactic plane.

\section{Summary}
We have found a non-thermal radio source at the location of the extended gamma-ray source 4FGL~J\,$1723.5-0501$e (FHES~J\,$1723.5-0501$), which we identify as a new SNR, \source. The size of its radio shell is  $51'\times45'$. The gamma rays show a hard spectrum consistent with a leptonic (IC) model. This model predicts synchrotron emission by the corresponding particle distribution whose spectrum is consistent with our measured radio spectrum. Such a scenario for the origin of the GeV emission is also expected in a low-density environment. This would be consistent with the location of the SNR, which is found outside the Galactic plane, where low ambient densities are expected. Comparing the radio and gamma-ray features of \source with those of the known SNR populations, we obtained a distance to the object in the range 1.4--3.5\,kpc. An SNR age of the order of 10\,kyr is compatible with the radio and GeV features, but an older or younger SNR cannot be ruled out. More multiwavelength observations of this object should be carried out.

\section*{Acknowledgements}

We thank the anonymous referee for useful comments that helped improved this work. MA and SQ received funding from Universidad de Costa Rica grant B8267 and acknowledge the use of computational resources from CICIMA-UCR. NHW is supported by an Australian Research Council Future Fellowship (project number FT190100231) funded by the Australian Government. We have made use of the ROSAT Data Archive of the MaxPlanck-Institut f{\"u}r extraterrestrische Physik (MPE) at Garching, Germany.

\section*{Data availability.} The derived data generated in this research will be shared on request to the corresponding author.

\section{Appendix}

\bibliographystyle{mnras}
\bibliography{newSNRG17_v3}

\begin{thebibliography}{}
\makeatletter
\relax
\def\mn@urlcharsother{\let\do\@makeother \do\$\do\&\do\#\do\^\do\_\do\%\do\~}
\def\mn@doi{\begingroup\mn@urlcharsother \@ifnextchar [ {\mn@doi@}
  {\mn@doi@[]}}
\def\mn@doi@[#1]#2{\def\@tempa{#1}\ifx\@tempa\@empty \href
  {http://dx.doi.org/#2} {doi:#2}\else \href {http://dx.doi.org/#2} {#1}\fi
  \endgroup}
\def\mn@eprint#1#2{\mn@eprint@#1:#2::\@nil}
\def\mn@eprint@arXiv#1{\href {http://arxiv.org/abs/#1} {{\tt arXiv:#1}}}
\def\mn@eprint@dblp#1{\href {http://dblp.uni-trier.de/rec/bibtex/#1.xml}
  {dblp:#1}}
\def\mn@eprint@#1:#2:#3:#4\@nil{\def\@tempa {#1}\def\@tempb {#2}\def\@tempc
  {#3}\ifx \@tempc \@empty \let \@tempc \@tempb \let \@tempb \@tempa \fi \ifx
  \@tempb \@empty \def\@tempb {arXiv}\fi \@ifundefined
  {mn@eprint@\@tempb}{\@tempb:\@tempc}{\expandafter \expandafter \csname
  mn@eprint@\@tempb\endcsname \expandafter{\@tempc}}}

\bibitem[\protect\citeauthoryear{{Abdo} et~al.,}{{Abdo}
  et~al.}{2011}]{2011ApJ...734...28A}
{Abdo} A.~A.,  et~al., 2011, \mn@doi [\apj] {10.1088/0004-637X/734/1/28}, \href
  {https://ui.adsabs.harvard.edu/abs/2011ApJ...734...28A} {734, 28}

\bibitem[\protect\citeauthoryear{Abdo et~al.,}{Abdo et~al.}{2013}]{Abdo_2013}
Abdo A.~A.,  et~al., 2013, \mn@doi [The Astrophysical Journal Supplement
  Series] {10.1088/0067-0049/208/2/17}, 208, 17

\bibitem[\protect\citeauthoryear{{Abdollahi} et~al.,}{{Abdollahi}
  et~al.}{2020}]{2020ApJS..247...33A}
{Abdollahi} S.,  et~al., 2020, \mn@doi [\apjs] {10.3847/1538-4365/ab6bcb},
  \href {https://ui.adsabs.harvard.edu/abs/2020ApJS..247...33A} {247, 33}

\bibitem[\protect\citeauthoryear{{Acero} et~al.,}{{Acero}
  et~al.}{2016}]{2016ApJS..224....8A}
{Acero} F.,  et~al., 2016, \mn@doi [\apjs] {10.3847/0067-0049/224/1/8}, \href
  {https://ui.adsabs.harvard.edu/abs/2016ApJS..224....8A} {224, 8}

\bibitem[\protect\citeauthoryear{Ackermann et~al.,}{Ackermann
  et~al.}{2018}]{Ackermann_2018}
Ackermann M.,  et~al., 2018, \mn@doi [The Astrophysical Journal Supplement
  Series] {10.3847/1538-4365/aacdf7}, 237, 32

\bibitem[\protect\citeauthoryear{{Ajello} et~al.,}{{Ajello}
  et~al.}{2016}]{2016ApJ...819...98A}
{Ajello} M.,  et~al., 2016, \mn@doi [\apj] {10.3847/0004-637X/819/2/98}, \href
  {https://ui.adsabs.harvard.edu/abs/2016ApJ...819...98A} {819, 98}

\bibitem[\protect\citeauthoryear{{Akaike}}{{Akaike}}{1974}]{1974ITAC...19..716A}
{Akaike} H.,  1974, IEEE Transactions on Automatic Control, \href
  {http://adsabs.harvard.edu/abs/1974ITAC...19..716A} {19, 716}

\bibitem[\protect\citeauthoryear{{Albert} et~al.,}{{Albert}
  et~al.}{2020a}]{2020ApJ...903L..14A}
{Albert} A.,  et~al., 2020a, \mn@doi [\apjl] {10.3847/2041-8213/abbfae}, \href
  {https://ui.adsabs.harvard.edu/abs/2020ApJ...903L..14A} {903, L14}

\bibitem[\protect\citeauthoryear{{Albert} et~al.,}{{Albert}
  et~al.}{2020b}]{2020ApJ...905...76A}
{Albert} A.,  et~al., 2020b, \mn@doi [\apj] {10.3847/1538-4357/abc2d8}, \href
  {https://ui.adsabs.harvard.edu/abs/2020ApJ...905...76A} {905, 76}

\bibitem[\protect\citeauthoryear{{Anderson} et~al.,}{{Anderson}
  et~al.}{2017}]{2017A&A...605A..58A}
{Anderson} L.~D.,  et~al., 2017, \mn@doi [\aap] {10.1051/0004-6361/201731019},
  \href {https://ui.adsabs.harvard.edu/abs/2017A&A...605A..58A} {605, A58}

\bibitem[\protect\citeauthoryear{{Araya}}{{Araya}}{2018}]{2018MNRAS.474..102A}
{Araya} M.,  2018, \mn@doi [\mnras] {10.1093/mnras/stx2779}, \href
  {https://ui.adsabs.harvard.edu/abs/2018MNRAS.474..102A} {474, 102}

\bibitem[\protect\citeauthoryear{{Araya}}{{Araya}}{2020}]{2020MNRAS.492.5980A}
{Araya} M.,  2020, \mn@doi [\mnras] {10.1093/mnras/staa244}, \href
  {https://ui.adsabs.harvard.edu/abs/2020MNRAS.492.5980A} {492, 5980}

\bibitem[\protect\citeauthoryear{{Atwood} et~al.,}{{Atwood}
  et~al.}{2009}]{2009ApJ...697.1071A}
{Atwood} W.~B.,  et~al., 2009, \mn@doi [\apj] {10.1088/0004-637X/697/2/1071},
  \href {https://ui.adsabs.harvard.edu/abs/2009ApJ...697.1071A} {697, 1071}

\bibitem[\protect\citeauthoryear{{Ballet}, {Burnett}, {Digel}  \&
  {Lott}}{{Ballet} et~al.}{2020}]{2020arXiv200511208B}
{Ballet} J.,  {Burnett} T.~H.,  {Digel} S.~W.,   {Lott} B.,  2020, arXiv
  e-prints, \href {https://ui.adsabs.harvard.edu/abs/2020arXiv200511208B} {p.
  arXiv:2005.11208}

\bibitem[\protect\citeauthoryear{{Becker}, {Hurley-Walker}, {Weinberger},
  {Nicastro}, {Mayer}, {Merloni}  \& {Sanders}}{{Becker}
  et~al.}{2021}]{2021A&A...648A..30B}
{Becker} W.,  {Hurley-Walker} N.,  {Weinberger} C.,  {Nicastro} L.,  {Mayer}
  M.~G.~F.,  {Merloni} A.,   {Sanders} J.,  2021, \mn@doi [\aap]
  {10.1051/0004-6361/202040156}, \href
  {https://ui.adsabs.harvard.edu/abs/2021A&A...648A..30B} {648, A30}

\bibitem[\protect\citeauthoryear{{Blasi}}{{Blasi}}{2013}]{2013A&ARv..21...70B}
{Blasi} P.,  2013, \mn@doi [\aapr] {10.1007/s00159-013-0070-7}, \href
  {https://ui.adsabs.harvard.edu/abs/2013A&ARv..21...70B} {21, 70}

\bibitem[\protect\citeauthoryear{{Brose}, {Pohl}  \& {Sushch}}{{Brose}
  et~al.}{2021}]{2021arXiv210810773B}
{Brose} R.,  {Pohl} M.,   {Sushch} I.,  2021, arXiv e-prints, \href
  {https://ui.adsabs.harvard.edu/abs/2021arXiv210810773B} {p. arXiv:2108.10773}

\bibitem[\protect\citeauthoryear{{Bruzewski}, {Schinzel}, {Taylor}  \&
  {Petrov}}{{Bruzewski} et~al.}{2021}]{2021ApJ...914...42B}
{Bruzewski} S.,  {Schinzel} F.~K.,  {Taylor} G.~B.,   {Petrov} L.,  2021,
  \mn@doi [\apj] {10.3847/1538-4357/abf73b}, \href
  {https://ui.adsabs.harvard.edu/abs/2021ApJ...914...42B} {914, 42}

\bibitem[\protect\citeauthoryear{{Calabretta}, {Staveley-Smith}  \&
  {Barnes}}{{Calabretta} et~al.}{2014}]{2014PASA...31....7C}
{Calabretta} M.~R.,  {Staveley-Smith} L.,   {Barnes} D.~G.,  2014, \mn@doi
  [\pasa] {10.1017/pasa.2013.36}, \href
  {https://ui.adsabs.harvard.edu/abs/2014PASA...31....7C} {31, e007}

\bibitem[\protect\citeauthoryear{{Carretti} et~al.,}{{Carretti}
  et~al.}{2019}]{2019MNRAS.489.2330C}
{Carretti} E.,  et~al., 2019, \mn@doi [\mnras] {10.1093/mnras/stz806}, \href
  {https://ui.adsabs.harvard.edu/abs/2019MNRAS.489.2330C} {489, 2330}

\bibitem[\protect\citeauthoryear{{Case} \& {Bhattacharya}}{{Case} \&
  {Bhattacharya}}{1998}]{1998ApJ...504..761C}
{Case} G.~L.,  {Bhattacharya} D.,  1998, \mn@doi [\apj] {10.1086/306089}, \href
  {https://ui.adsabs.harvard.edu/abs/1998ApJ...504..761C} {504, 761}

\bibitem[\protect\citeauthoryear{{Condon}, {Cotton}, {Greisen}, {Yin},
  {Perley}, {Taylor}  \& {Broderick}}{{Condon}
  et~al.}{1998}]{1998AJ....115.1693C}
{Condon} J.~J.,  {Cotton} W.~D.,  {Greisen} E.~W.,  {Yin} Q.~F.,  {Perley}
  R.~A.,  {Taylor} G.~B.,   {Broderick} J.~J.,  1998, \mn@doi [\aj]
  {10.1086/300337}, \href
  {https://ui.adsabs.harvard.edu/abs/1998AJ....115.1693C} {115, 1693}

\bibitem[\protect\citeauthoryear{{Cotton}}{{Cotton}}{2017}]{2017PASP..129i4501C}
{Cotton} W.~D.,  2017, \mn@doi [\pasp] {10.1088/1538-3873/aa793f}, \href
  {https://ui.adsabs.harvard.edu/abs/2017PASP..129i4501C} {129, 094501}

\bibitem[\protect\citeauthoryear{{Dai}, {Wang}, {Vadakkumthani}  \&
  {Xing}}{{Dai} et~al.}{2016}]{2016RAA....16...97D}
{Dai} X.-J.,  {Wang} Z.-X.,  {Vadakkumthani} J.,   {Xing} Y.,  2016, \mn@doi
  [Research in Astronomy and Astrophysics] {10.1088/1674-4527/16/6/097}, \href
  {https://ui.adsabs.harvard.edu/abs/2016RAA....16...97D} {16, 97}

\bibitem[\protect\citeauthoryear{{Devin}, {Lemoine-Goumard}, {Grondin},
  {Castro}, {Ballet}, {Cohen}  \& {Hewitt}}{{Devin}
  et~al.}{2020}]{2020A&A...643A..28D}
{Devin} J.,  {Lemoine-Goumard} M.,  {Grondin} M.~H.,  {Castro} D.,  {Ballet}
  J.,  {Cohen} J.,   {Hewitt} J.~W.,  2020, \mn@doi [\aap]
  {10.1051/0004-6361/202038503}, \href
  {https://ui.adsabs.harvard.edu/abs/2020A&A...643A..28D} {643, A28}

\bibitem[\protect\citeauthoryear{{Dubner} \& {Giacani}}{{Dubner} \&
  {Giacani}}{2015}]{2015A&ARv..23....3D}
{Dubner} G.,  {Giacani} E.,  2015, \mn@doi [\aapr] {10.1007/s00159-015-0083-5},
  \href {https://ui.adsabs.harvard.edu/abs/2015A&ARv..23....3D} {23, 3}

\bibitem[\protect\citeauthoryear{{Gabici} \& {Aharonian}}{{Gabici} \&
  {Aharonian}}{2014}]{2014MNRAS.445L..70G}
{Gabici} S.,  {Aharonian} F.~A.,  2014, \mn@doi [\mnras]
  {10.1093/mnrasl/slu132}, \href
  {https://ui.adsabs.harvard.edu/abs/2014MNRAS.445L..70G} {445, L70}

\bibitem[\protect\citeauthoryear{{Gao} \& {Han}}{{Gao} \&
  {Han}}{2014}]{2014A&A...567A..59G}
{Gao} X.~Y.,  {Han} J.~L.,  2014, \mn@doi [\aap] {10.1051/0004-6361/201424128},
  \href {https://ui.adsabs.harvard.edu/abs/2014A&A...567A..59G} {567, A59}

\bibitem[\protect\citeauthoryear{{Green}}{{Green}}{2014}]{2014BASI...42...47G}
{Green} D.~A.,  2014, Bulletin of the Astronomical Society of India, \href
  {https://ui.adsabs.harvard.edu/abs/2014BASI...42...47G} {42, 47}

\bibitem[\protect\citeauthoryear{{Green}}{{Green}}{2019}]{2019JApA...40...36G}
{Green} D.~A.,  2019, \mn@doi [Journal of Astrophysics and Astronomy]
  {10.1007/s12036-019-9601-6}, \href
  {https://ui.adsabs.harvard.edu/abs/2019JApA...40...36G} {40, 36}

\bibitem[\protect\citeauthoryear{{H.~E.~S.~S. Collaboration}
  et~al.,}{{H.~E.~S.~S. Collaboration} et~al.}{2018}]{2018A&A...612A...1H}
{H.~E.~S.~S. Collaboration} et~al., 2018, \mn@doi [\aap]
  {10.1051/0004-6361/201732098}, \href
  {https://ui.adsabs.harvard.edu/abs/2018A&A...612A...1H} {612, A1}

\bibitem[\protect\citeauthoryear{{Hurley-Walker} et~al.,}{{Hurley-Walker}
  et~al.}{2019}]{2019PASA...36...48H}
{Hurley-Walker} N.,  et~al., 2019, \mn@doi [\pasa] {10.1017/pasa.2019.33},
  \href {https://ui.adsabs.harvard.edu/abs/2019PASA...36...48H} {36, e048}

\bibitem[\protect\citeauthoryear{{Lande} et~al.,}{{Lande}
  et~al.}{2012}]{2012ApJ...756....5L}
{Lande} J.,  et~al., 2012, \mn@doi [\apj] {10.1088/0004-637X/756/1/5}, \href
  {https://ui.adsabs.harvard.edu/abs/2012ApJ...756....5L} {756, 5}

\bibitem[\protect\citeauthoryear{{Leahy} \& {Williams}}{{Leahy} \&
  {Williams}}{2017}]{2017AJ....153..239L}
{Leahy} D.~A.,  {Williams} J.~E.,  2017, \mn@doi [\aj]
  {10.3847/1538-3881/aa6af6}, \href
  {https://ui.adsabs.harvard.edu/abs/2017AJ....153..239L} {153, 239}

\bibitem[\protect\citeauthoryear{{Lorimer}, {Bailes}  \& {Harrison}}{{Lorimer}
  et~al.}{1997}]{1997MNRAS.289..592L}
{Lorimer} D.~R.,  {Bailes} M.,   {Harrison} P.~A.,  1997, \mn@doi [\mnras]
  {10.1093/mnras/289.3.592}, \href
  {https://ui.adsabs.harvard.edu/abs/1997MNRAS.289..592L} {289, 592}

\bibitem[\protect\citeauthoryear{{Manchester}, {Hobbs}, {Teoh}  \&
  {Hobbs}}{{Manchester} et~al.}{2005}]{2005AJ....129.1993M}
{Manchester} R.~N.,  {Hobbs} G.~B.,  {Teoh} A.,   {Hobbs} M.,  2005, \mn@doi
  [\aj] {10.1086/428488}, \href
  {https://ui.adsabs.harvard.edu/abs/2005AJ....129.1993M} {129, 1993}

\bibitem[\protect\citeauthoryear{{Mattox} et~al.,}{{Mattox}
  et~al.}{1996}]{1996ApJ...461..396M}
{Mattox} J.~R.,  et~al., 1996, \mn@doi [\apj] {10.1086/177068}, \href
  {http://adsabs.harvard.edu/abs/1996ApJ...461..396M} {461, 396}

\bibitem[\protect\citeauthoryear{{Ohira}, {Yamazaki}, {Kawanaka}  \&
  {Ioka}}{{Ohira} et~al.}{2012}]{2012MNRAS.427...91O}
{Ohira} Y.,  {Yamazaki} R.,  {Kawanaka} N.,   {Ioka} K.,  2012, \mn@doi
  [\mnras] {10.1111/j.1365-2966.2012.21908.x}, \href
  {https://ui.adsabs.harvard.edu/abs/2012MNRAS.427...91O} {427, 91}

\bibitem[\protect\citeauthoryear{{Reynolds}}{{Reynolds}}{2008}]{2008ARA&A..46...89R}
{Reynolds} S.~P.,  2008, \mn@doi [\araa]
  {10.1146/annurev.astro.46.060407.145237}, \href
  {https://ui.adsabs.harvard.edu/abs/2008ARA&A..46...89R} {46, 89}

\bibitem[\protect\citeauthoryear{{Saz Parkinson}, {Xu}, {Yu}, {Salvetti},
  {Marelli}  \& {Falcone}}{{Saz Parkinson} et~al.}{2016}]{2016ApJ...820....8S}
{Saz Parkinson} P.~M.,  {Xu} H.,  {Yu} P.~L.~H.,  {Salvetti} D.,  {Marelli} M.,
    {Falcone} A.~D.,  2016, \mn@doi [\apj] {10.3847/0004-637X/820/1/8}, \href
  {https://ui.adsabs.harvard.edu/abs/2016ApJ...820....8S} {820, 8}

\bibitem[\protect\citeauthoryear{{Uro{\v{s}}evi{\'c}}}{{Uro{\v{s}}evi{\'c}}}{2014}]{2014Ap&SS.354..541U}
{Uro{\v{s}}evi{\'c}} D.,  2014, \mn@doi [\apss] {10.1007/s10509-014-2095-4},
  \href {https://ui.adsabs.harvard.edu/abs/2014Ap&SS.354..541U} {354, 541}

\bibitem[\protect\citeauthoryear{{Voges} et~al.,}{{Voges}
  et~al.}{1999}]{1999A&A...349..389V}
{Voges} W.,  et~al., 1999, \aap, \href
  {https://ui.adsabs.harvard.edu/abs/1999A&A...349..389V} {349, 389}

\bibitem[\protect\citeauthoryear{{Xing}, {Wang}, {Zhang}  \& {Chen}}{{Xing}
  et~al.}{2019}]{2019PASJ...71...77X}
{Xing} Y.,  {Wang} Z.,  {Zhang} X.,   {Chen} Y.,  2019, \mn@doi [\pasj]
  {10.1093/pasj/psz056}, \href
  {https://ui.adsabs.harvard.edu/abs/2019PASJ...71...77X} {71, 77}

\bibitem[\protect\citeauthoryear{{Yasuda} \& {Lee}}{{Yasuda} \&
  {Lee}}{2019}]{2019ApJ...876...27Y}
{Yasuda} H.,  {Lee} S.-H.,  2019, \mn@doi [\apj] {10.3847/1538-4357/ab13ab},
  \href {https://ui.adsabs.harvard.edu/abs/2019ApJ...876...27Y} {876, 27}

\bibitem[\protect\citeauthoryear{{Zabalza}}{{Zabalza}}{2015}]{naima}
{Zabalza} V.,  2015, Proc.~of International Cosmic Ray Conference 2015, \href
  {http://adsabs.harvard.edu/abs/2015arXiv150903319Z} {p.~922}

\bibitem[\protect\citeauthoryear{{de Gasperin}, {Intema}  \& {Frail}}{{de
  Gasperin} et~al.}{2018}]{2018MNRAS.474.5008D}
{de Gasperin} F.,  {Intema} H.~T.,   {Frail} D.~A.,  2018, \mn@doi [\mnras]
  {10.1093/mnras/stx3125}, \href
  {https://ui.adsabs.harvard.edu/abs/2018MNRAS.474.5008D} {474, 5008}

\makeatother
\end{thebibliography}

% Don't change these lines
\bsp    % typesetting comment
\label{lastpage}

\end{document}